\documentstyle[epsf]{l-aa}
\thesaurus{ 06(02.16.1 ; 02.16.2 ; 08.16.6 ; 08.16.7 PSR B0144+59 ;
08.16.7 PSR B1737$-$30 ; 08.16.7 PSR B1913+10)
}

\begin{document}

\title{A transition from linear to circular polarization in pulsar radio emission}

\author{ A. ~von Hoensbroech \inst{1,2} and H. ~Lesch \inst{2}}

\offprints {A.~von~Hoensbroech (avh@mpifr-bonn.mpg.de)}

\institute{Max-Planck-Institut f\"ur Radioastronomie,
 Auf dem H\"ugel 69, D-53121 Bonn, Germany.
\and
Institut f\"ur Astronomie und Astrophysik der
 Ludwig-Maximilian-Universit\"at M\"unchen, Scheinerstr. 1, D-81679
 M\"unchen, Germany.}

\date{Received date ; accepted date}

\maketitle

\markboth{~von Hoensbroech \& Lesch: A transition from linear to
circular ...}{}

\begin{abstract}

At present three pulsars are known which clearly show a strongly
increasing degree of circular polarization with frequency. As this is
accompanied by a smooth decrease of linear polarization, we
investigate, if this observation can be explained through a
propagation mode transition from linear to circular polarization.
Using a previously published model we find that the rate of change of
the polarization types with frequency is well consistent with the
theory.

The small number of only three objects does of course not allow
to draw a general conclusion for the whole sample of pulsars. However
we show that the very unusual behavior of these three objects can
be well explained with this model and we discuss, why only so few
such objects are presently known.

\end{abstract}


\section{Introduction}

Whether the observed polarimetric features of pulsar radio emission
are intrinsic properties of the emission mechanism or if they are a
result of propagational effects is one of the open questions for the
understanding of the polarization of radio pulsars. Early
observations of the `S'-type swing of the polarization position angle
through the pulse profile of the Vela pulsar led to an geometrical
interpretation by Radhakrishnan \& Cooke (1969). In their model the
swing reflects the geometry of the magnetic field lines projected on
the plane of the sky. In some cases the expected swing is matched so
well that a non-geometrical origin for the position angle seems rather
improbable. Ruderman \& Sutherland (1975) gave an explanation for this
phenomenon as an intrinsic property of emission by coherent curvature
radiation. The particles experience an acceleration perpendicular to
their motion in the plane of curvature of the field line. The
resulting radiation has its electric field vector within this plane,
thus producing the geometrical signature.

Alternatively, it was proposed that the observed polarization
originates from propagational effects or is at least influenced by
them.  Various authors calculated the properties of the propagational
modes in a pulsar magnetosphere (e.g. Allan \& Melrose 1982, Barnard
\& Arons 1986, Lyutikov 1998 and references therein). It is widely
agreed among those authors, that two independent, generally elliptical
polarization modes propagate, which are oriented parallel and
perpendicular to the plane of curvature of the magnetic field line.
The observed polarization then corresponds to the shape of these modes
at the distance from the neutron star, where the radiation decouples
from the plasma. This distance is called the {\it polarization
limiting radius} (hereafter PLR) and does not necessarily coincide
with the place of emission. The magnitude of the PLR has been a
subject of theoretical debates. Barnard (1986) and Beskin et
al. (1993) place the PLR at the cyclotron resonance. Melrose (1979)
defines a coupling ratio $Q=1/(\Delta k L)$ where $\Delta k$ denotes
the difference in wavenumber of the two modes and $L$ is a
characteristic length scale. The PLR is at the position where $Q\simeq
1$.

In a previous paper (von Hoensbroech et al. 1998b) we have shown that
many of the complex variety of radio pulsar polarization states can be
understood qualitatively if some propagational effects in the pulsar
magnetosphere are taken into account. This was done by using a simple
approximation for the properties of natural polarization modes as they
propagate through the magnetosphere. The model is based on the
following assumptions: {\it 1.} The temperature of the background
plasma is assumed to be zero (distribution function
$f(\gamma)=\delta(\gamma-\gamma_{\rm bg})$, hereafter $\gamma_{\rm
bg}$ is the $\gamma$-factor of the background plasma).  {\it 2.}
Furthermore no strong pair production in a sense, that the
$e^-e^+$-plasma is not approximately neutral, is assumed.  Considering
all relevant angles under which the propagating wave and the magnetic
field lines intersect, the shape of the natural polarization modes can
be calculated at any given point between the emission height and the
light cylinder radius (hereafter $R_{\rm LC}$, $R_{\rm
LC}=cP/2\pi$). {\it 3.} Assuming a PLR and a certain Lorentz-factor
for the background plasma, the following qualitative statements can be
made: {\it a.} The polarization modes propagate independently with
their position angle parallel and orthogonal to the local plane of
field line curvature, {\it b.} the radiation depolarizes towards high
frequencies, {\it c.}  the degree of polarization correlates with the
pulsars loss of rotational energy $\dot E$ and {\it d.} the
polarization smoothly changes from linear to circular with increasing
frequency.

This last point was met by recent observations at the relatively high
radio frequencies of $\nu_{\rm obs}=4.9$ GHz (von Hoensbroech et
al. 1998a) as it was already suspected by early measurements of Morris
et al. (1981). For a few pulsars the degree of circular polarization
increases strongly with frequency. For three of those objects, the
circular polarization reaches values of more than 50 \%, even
surpassing the linear polarization. This rather unusual behavior
appears smooth and is thus likely to be systematic, as an inspection
of lower frequency data showed. We therefore use the data of those
three clear examples to quantitatively compare the rate of change from
linear to circular to the predicted one in statement {\it d}.

\section{Analysis and Results}

\begin{figure}[t]
\epsfxsize=9cm
\hspace{1.2cm}\epsfbox{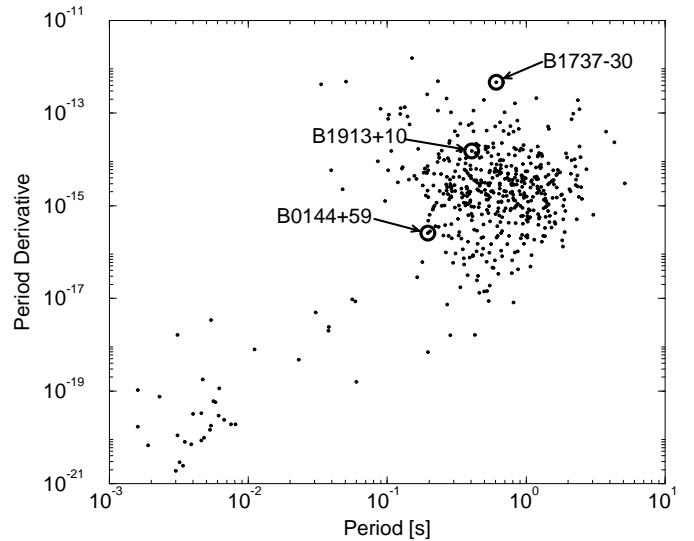}
\caption[h]{Position of the three pulsars in the $P-\dot P$--diagram
for which the change from linear to circular polarization has been
observed. They do not form a coherent group within the
pulsar sample. One of them is very young, one is rather average and
the third one is a relatively old object.}
\end{figure}

\begin{table}[t]
\caption[]{Polarization data and references for all three pulsars.}
\begin{tabular}{llllll}
\noalign{\smallskip}
\hline
\noalign{\smallskip}
Pulsar&Freq. [MHz]&$\Pi_{\rm L}$[\%]&$\Pi_{\rm C}[\%]$&$R$&Ref.\footnotemark\\
\noalign{\smallskip}
\hline
\hline
\noalign{\smallskip}
B0144+59& 610  & $29 \pm 2$ & $11 \pm 3 $&$ 2.64 \pm 0.74 $&[1]\\
	& 1408 & $23 \pm 1$ & $24 \pm 2 $&$ 0.96 \pm 0.09 $&[1]\\
	& 1642 & $25 \pm 6$ & $26 \pm 6 $&$ 0.96 \pm 0.32 $&[1]\\
	& 2695 & $16 \pm 3$ & $50 \pm 4 $&$ 0.32 \pm 0.07 $&[4]\\
	& 4850 & $12 \pm 1$ & $50 \pm 2 $&$ 0.24 \pm 0.02 $&[2]\\
\noalign{\smallskip}
\hline
\noalign{\smallskip}
B1737$-$30& 610  &$22\pm 1$   &$ 20\pm 1$  &$ 1.10\pm 0.07 $&[1]\\
	& 660  &$43\pm 3$   &$ 17\pm 4$  &$ 2.53\pm 0.62 $&[3]\\
	& 925  &$73\pm 2$   &$ 21\pm 2$  &$ 3.48\pm 0.34 $&[1]\\
	& 1408 &$80\pm 1$   &$ 38\pm 1$  &$ 2.11\pm 0.06 $&[1]\\
	& 1642 &$71\pm 1$   &$ 45\pm 1$  &$ 1.58\pm 0.04 $&[1]\\
	& 4750 &$53\pm 7$   &$ 59\pm 5$  &$ 0.90\pm 0.14 $&[2]\\
\noalign{\smallskip}
\hline
\noalign{\smallskip}
B1913+10& 408  &$18\pm 2$   &$ 13\pm 3$  &$ 1.38\pm 0.35 $&[1]\\
	& 610  &$38\pm 2$   &$ 22\pm 2$  &$ 1.73\pm 0.18 $&[1]\\
	& 925  &$77\pm 4$   &$ 35\pm 3$  &$ 2.20\pm 0.22 $&[1]\\
	& 1408 &$50\pm 1$   &$ 42\pm 1$  &$ 1.19\pm 0.04 $&[1]\\
	& 1642 &$49\pm 2$   &$ 46\pm 3$  &$ 1.07\pm 0.08 $&[1]\\
	& 4850 &$21\pm 4$   &$ 55\pm 5$  &$ 0.38\pm 0.08 $&[2]\\
\noalign{\smallskip}
\hline
\end{tabular}
\end{table}

\footnotetext{Reference-code: [1] Gould \& Lyne (1997), [2] von
Hoensbroech et al. (1998a), [3] Qiao et al. (1995), [4] unpublished
Effelsberg data.}

Presently we have data for three pulsars, where a clear change from linear
to circular polarization towards high frequencies is observed
(see Fig. 5 in von Hoensbroech et al. 1998b). It is obvious from Fig. 1
that the three objects have very different rotational periods $P$ and
period time derivatives $\dot P$.
Hence they do not form an isolated group with respect to their
rotational parameters.
Since the change from linear to circular polarization is superposed by
general depolarization effects,
which affect both types of polarization likewise, the degrees of
polarization cannot be compared directly with the theory. Therefore we
choose the ratio
\begin{equation}
R:=\frac{\Pi_L}{\Pi_C}
\end{equation}
between the degrees of linear ($\Pi_L$) and the circular ($\Pi_C$)
polarization as an depolarization independent parameter. $R$ and its
statistical error $\Delta R$ can easily be calculated from the data.

The polarization data we used for this analysis were accessed
through the online EPN-database ({\it
http://www.mpifr-bonn.mpg.de/pulsar/data/}).
The calibrated data were extracted in the
EPN-format (Lorimer et al. 1998). The degrees of polarization were
calculated using the same routine for all profiles.
All relevant parameters and references are
listed in Table 1.

The theoretical functional dependence of $R$ can be derived from
equation (25) in (von Hoensbroech et al. 1998b). Applying a
Taylor-expansion in inverse frequency $\nu^{-1}$ to the first order yields

\begin{eqnarray}\label{Req}
R(\nu)&=&\frac{\Omega_{e,{\rm PLR}}}{\gamma_{\rm bg}}
\frac{\sin^2\theta_{\rm PLR}}{4\pi(\cos\theta_{\rm PLR}-\beta)
(1-\beta\cos\theta_{\rm PLR})^2}
\cdot\nu^{-1}\\
&&+O\left(\nu^{-3}\right)\ .
\end{eqnarray}

Here $\Omega_{e,{\rm PLR}}$ is the local electron gyro frequency at the
PLR, $\theta_{\rm PLR}$ the angle between the propagating wave and the
direction of the local magnetic field at the PLR and
$\beta=v/c=\sqrt{1-1/\gamma_{\rm bg}^2}$ the streaming velocity of the
plasma.  Higher order terms can be neglected as long as the wave
frequency is different from $\sim\Omega_e/2\pi$.

Obviously, a $\nu^{-1}$--frequency dependence of $R$ is required by
the theory. A comparison of the observed frequency dependence of $R$
with the predicted one is therefore a strong test for the theory.
Figs. 2 -- 4 show a comparison of the data
points and the theoretical curve (dashed line) for $R$.

There are a couple of parameters which indirectly enter
Eq. (\ref{Req}) as scaling factors. $\Omega_{e,{\rm PLR}}$ is
proportional to the local magnetic field strength $B(PLR)$. This value
again depends on basic pulsar parameters such as the period, its time
derivative and, if known, on the inclination angle between the
rotation- and the magnetic dipole axis. Furthermore the angle
$\theta_{\rm PLR}$ between the propagating wave and the local magnetic
field depends on the chosen field line and the assumed emission
height. For the background plasma Lorentz factor we made the
assumption $\gamma_{\rm bg}=200$. Finally we chose the PLR at 20 \% of
$R_{\rm LC}$. As the values of the PLR and the emission height (2 \%
of $R_{\rm LC}$ are given as fractions of $R_{\rm LC}$, their absolute
values depend on the period. The combination of $\gamma_{\rm bg}$ and
PLR was chosen without restriction of generality as various other
combinations yield the same result (see Fig. 5).

However, apart from the known intrinsic pulsar parameters $P$ and
$\dot P$, the {\it same} set of parameters was used for all three
pulsars. Please note that these parameters only enter Eq. (\ref{Req})
as scaling factors, yielding a parallel displacement of the
function. Hence they determine the frequency range where the
transition from linear to circular polarization takes place, but {\it
not the functional dependence} $\sim \nu^{-1}$.

\begin{figure}[t]
\epsfxsize=8.5cm
\epsfbox{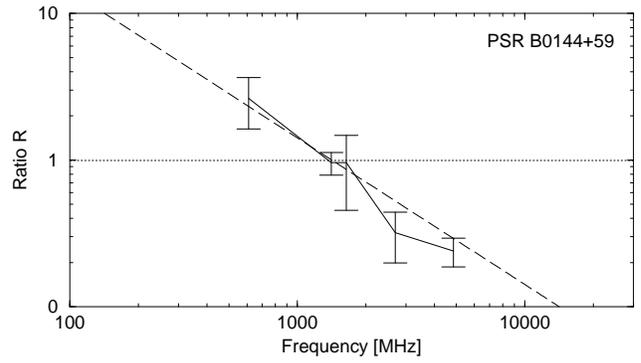}
\caption[h]{PSR B0144+59, $R=\Pi_L/\Pi_C$ versus frequency. The
theoretical change of $R$ (dashed line) is compared to measured
data. The horizontal dotted line corresponds to $R=1$, i.e. linear and
circular polarization are of equal strength (here $R=1$ at $\nu=1.4$ GHz).
Parameters for theoretical line: $\gamma_{\rm bg}=200, {\rm PLR}=20\%$
of $R_{\rm LC}$.
}
\end{figure}
\begin{figure}[t]
\epsfxsize=8.5cm
\epsfbox{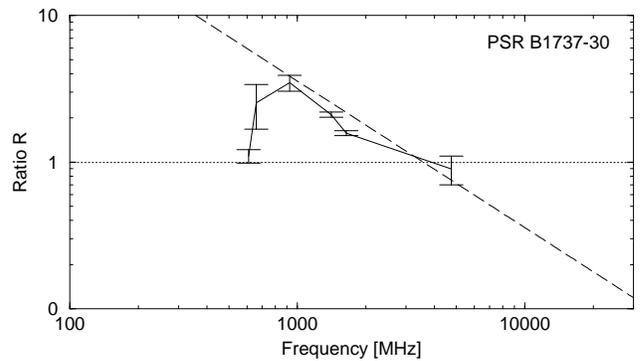}
\caption[h]{PSR B1737$-$30, see caption of Fig. 2 and text for details
(here $R=1$ at $\nu=3.5$ GHz). See text about the two ``outriders'.
Parameters for theoretical line: $\gamma_{\rm bg}=200, {\rm PLR}=20\%$
of $R_{\rm LC}$.
}
\end{figure}
\begin{figure}[t]
\epsfxsize=8.5cm
\epsfbox{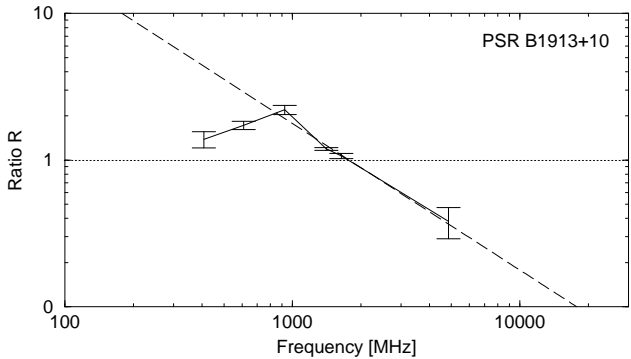}
\caption[h]{PSR B1913+10, see caption of Fig. 2 and text for
details (here $R=1$ at $\nu=1.8 $ GHz).
Parameters for theoretical line: $\gamma_{\rm bg}=200, {\rm PLR}=20\%$

}
\end{figure}
\begin{figure}[t]
\epsfxsize=9cm
\epsfbox{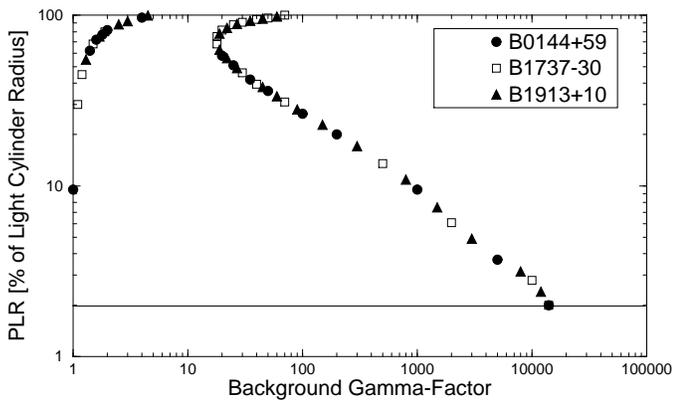}
\caption[h]{Possible combinations of PLR and $\gamma_{\rm bg}$ which
yield consistent results for the prefactor in Eq. (\ref{Req}) which
fit to the observed data of each pulsar. A variation of this factor
results into a parallel displacement of the theoretical curves in
Figs. 2-4.
The solid line below corresponds to the assumed emission height at 2
\% of the $R_{\rm LC}$ as the minimum PLR.
}
\end{figure}

\subsection{PSR B0144+59}
This pulsar is the first one in which we found the effect of increasing
circular polarization to high frequencies. Its rotational parameters
are $P=196$ ms and $\dot P=0.3\times10^{-15}$, yielding to a weak
surface magnetic field of `only' $B_0=2.3\times 10^7$ T, an average
value for $\dot E=1.3\times10^{26}$ W and the characteristic age
$\tau=1.2\times10^7$ yrs.
Reasonable data in full polarization was available between 610 MHz and
4.85 GHz.

Fig. 2 shows the measured values and the theoretical curve for the
change of $R$ with frequency.

\subsection{PSR B1737$-$30}\label{S1737}
PSR B1737$-$30 has a spin period of $P=607$ ms and a period derivative
of $\dot P=466\times10^{-15}$. The resulting spin down energy loss
$\dot E=8.3\times 10^{27} W$ places it amongst the top 10\% of the pulsar
sample. The very low characteristic age $\tau=2\times10^4$ yrs and the
very high surface magnetic field $B_0=1.7\times 10^9$ T make this one an
extreme object. Note that in terms of the surface magnetic
field, this object is at the opposite end of the ``normal'' pulsar sample
compared to PSR B0144+59.

The change of $R$ with frequency is shown in Fig. 3. Although the two
low frequency points do not fit the theoretical function, the
frequency change of $R$ smoothly follows this function at higher
frequencies.  Note that the two outrider profiles are significantly
affected by interstellar scattering which could also alter the
polarization state when emission from different pulse phases with
different polarization states are superposed (see also
Sect. \ref{outriders}).

\subsection{PSR B1913+10}
With a spin period $P=405$ ms and its temporal derivative $\dot
P=15\times10^{-15}$ this is an average pulsar. This is also reflected by
its parameters $\dot E=9\times10^{26}$ W, $B_0=2.5 \times 10^8$ T and
$\tau=4\times 10^5$ yrs.

Fig. 4 shows the measured values for $R$ and the theoretical
function . As for PSR B1737$-$30 the lower frequency points are too
small, but at higher frequencies the theoretical curve is matched
perfectly.  As for the previous pulsar, the two outrider profiles are
significantly scattered.

\subsection{Outriders}\label{outriders}

The systematic deviation of the low frequency points is certainly a
draw back of these observations. However they can be understood
through the following argument:  von Hoensbroech et al. (1998a) have
shown that the polarization properties in general are much less
systematic at low radio frequencies compared to higher ones. This
indicates that the polarization of pulsars undergo some sort of
randomization at low radio frequencies. This can
be caused either through intrinsic variations -- e.g. non constant
PLR at low frequencies -- or through additional propagation
effects in the highly magnetized medium close to the pulsar, which
mainly affect low radio frequencies.

\section{Discussion}

Based on a simple propagational scenario for natural polarization
modes in pulsar magnetospheres we were able to show that for the
available data set of three pulsars, theory and observations are in
very good agreement. In these objects the polarization changes
smoothly from linear to circular with increasing frequency. The
principal theoretical $\nu^{-1}$-functional dependence of of this
change, expressed in Eq. (\ref{Req}) is independent of all pulsar
parameters (as long as a mono-energetic background plasma is
concerned). The individual pulsar parameters enter Eq. (\ref{Req})
only as scaling factors, hence yielding a parallel displacement of the
function. Thus the comparison of the calculated and observed
functional dependence is a strong test for the theory. As shown in
Sect. 2 the rate of change of $R$ with frequency is consistent with
Eq. (\ref{Req}).

We note that for all three pulsars the {\it same} combinations of
$\gamma_{\rm bg}$ and PLR result into quantitative agreements between
the measured points and the theoretical curve, although these objects
are very different in their parameters $P$ and $\dot P$.  Obviously
not only the $\nu^{-1}$ dependence is in agreement with the
observation, but also the scaling of the function as it is defined
through the prefactor in Eq. (\ref{Req}) for these three pulsars.
This scaling coincidence is remarkable but does not necessarily mean,
that {\it all} pulsars have to scale in the same manner.

Of course the number of pulsars which clearly show this effect is too
small to make a statistically rigorous statement. This is especially
true as it cannot be predicted at which frequency this effect should
occur and hence for which other pulsars we should have observed it.
We emphasize that it is rather difficult to observe such a frequency
behavior of polarization since it requires a significant increase of
the circular polarization to result into some decrease of the linear
because they are added in quadrature.

Pulsars exhibit steep radio spectra, a fact which reduces the total
number of available profiles in full polarization to about 100 at 4.9
GHz with many of them at a fairly low signal-to-noise ratio. As
pulsars are also known to depolarize towards high frequencies, the
signal-to-noise-ratio of the polarized emission is often very low,
hardly allowing a credible determination of the ratio between linear
and circular polarization.  Nevertheless, it has been shown at low
frequencies that the linear polarization reduces with frequency while
such a trend is not evident for circular polarization (e.g. Han et
al. 1998). This result qualitatively fits into the same trend of
changing the ratio between the two polarizations.  Taking all points
mentioned above into account we are not surprised that only three
pulsars have been found yet which clearly show this effect and allow
an opportunity to test the theory.

In this paper we have used the frequency dependence of {\it
integrated} profiles. These integrated profiles of course differ
strongly from the individual pulses. It has been shown e.g. by
Stinebring et al. (1984), that these individual pulses are often
highly circularly polarized also at low frequencies. This is not
necessarily in contradiction to the ideas described in this paper as
the conditions within the magnetosphere are known to be highly
non-stationary and unstable.  The location at which the radiation
decouples from the plasma (PLR) might therefore vary strongly from
pulse to pulse.  However, the integrated profile represents an average
over these individual pulses, corresponding to an average PLR. A
crucial test whether the described effect is responsible for the
circular polarization would be the simultaneous observation of single
pulses in full polarization over large frequency intervals.

Summarizing, we may say that the observations of the three objects support the
theoretical approach that the polarization characteristics -- and
especially the high frequency occurrence of strong circular
polarization -- of at least these three pulsars are explainable
in terms of propagating natural wave modes in the magnetosphere.
Thus, it may be worth to follow that line of thought from the
theoretical point of view and to try to get some more high quality data
of the general polarization properties at low and high radio frequencies.

\acknowledgements
We thank A. Jessner, D. R. Lorimer, M. Kramer, T. Kunzl and 
R. Wielebinski for valuable comments on the manuscript and support
of this work.

\end{document}